\documentclass[reprint,superscriptaddress,amsmath,amssymb,aps,prd,notitlepage,longbibliography,floatfix,nofootinbib]{revtex4-1}

\usepackage{tensor}     
\usepackage{graphicx}   
\usepackage{subfigure}
\usepackage[
colorlinks=true,        
citecolor=blue,         
linkcolor=blue,         
urlcolor=blue           
]{hyperref}             
\usepackage{bm}         
\usepackage{xcolor}     
\usepackage{lipsum}
\usepackage{color}      
\usepackage[utf8]{inputenc} 
\usepackage[section]{placeins} 
\usepackage{multirow}
\usepackage[title]{appendix}
\usepackage{simpler-wick}
\usepackage{float}

\newcommand{\nc}{\newcommand*}

\nc{\al}{\alpha}
\nc{\s}{\sigma}
\nc{\kp}{\kappa}
\nc{\dt}{\delta}
\nc{\Dt}{\Delta}
\nc{\Ld}{\Lambda}
\nc{\p}{\partial}
\nc{\Gm}{\Gamma}
\nc{\om}{\omega}
\nc{\Om}{\Omega}
\nc{\rd}{\mathrm{d}}
\def\({\left(}
\def\){\right)}
\def\[{\left[}
\def\]{\right]}
\def\e{\begin{equation}}
\def\q{\end{equation}}
\def\m{\begin{eqnarray}}
\def\n{\end{eqnarray}}
\nc{\fnl}{F_\mathrm{nl}}
\nc{\gnl}{G_\mathrm{nl}}
\nc{\Eq}[1]{Eq.~\eqref{#1}}     
\nc{\Fig}[1]{Fig.~\ref{#1}}     
\nc{\Table}[1]{Table~\ref{#1}}  
\nc{\Sec}[1]{Sec.~\ref{#1}}     
\nc{\Msun}{M_\odot}             
\nc{\fpbh}{f_{\mathrm{PBH}}}    
\nc{\mpbh}{m_{\mathrm{PBH}}}    
\nc{\fpbhn}{f_{\mathrm{pbh0}}}    
\nc{\mR}{\mathcal{R}} 
\nc{\seq}{\sigma_{\mathrm{eq}}}
\nc{\ogw}{\Omega_{\mathrm{GW}}}
\nc{\gpcyr}{\mathrm{Gpc}^{-3}\,\mathrm{yr}^{-1}}
\nc{\lvc}{LIGO-Virgo-KAGRA} 
\nc{\SNR}{\mathrm{SNR}} 
\nc{\mmin}{{m_{\mathrm{min}}}}
\nc{\mmax}{{m_{\mathrm{max}}}}
\nc{\Mmin}{{M_{\mathrm{min}}}}
\nc{\fmin}{{f_{\mathrm{min}}}}
\nc{\VT}{\mathrm{VT}}
\nc{\rhoGW}{\rho_{\mathrm{GW}}}
\nc{\vth}{\vec{\theta}}
\nc{\vd}{\vec{d}}
\nc{\vla}{\vec{\lambda}}
\nc{\Nobs}{N_{\mathrm{obs}}}
\nc{\av}[1]{\langle #1 \rangle} 
\nc{\km}{\mathrm{km}}
\nc{\Mpc}{\mathrm{Mpc}}
\nc{\Tobs}{T_{\mathrm{obs}}}
\nc{\Ntemp}{N_{\mathrm{temp}}}
\nc{\fyr}{f_{\mathrm{yr}}}

\nc{\addref}{[\textcolor{red}{add ref}] } 
\nc{\eg}{\textit{e.g.~}}
\nc{\app}{\approx}
\nc{\hf}{\frac{1}{2}}
\nc{\discuss}{\textcolor{red}{Add discussion here!}}
\nc{\red}[1]{\textcolor{red}{#1}}
\nc{\hp}{h_+} 
\nc{\hc}{h_{\times}} 
\nc{\mbh}{M_{\rm BH}}
\nc{\mdisk}{M_{\mathrm{disk}}}
\nc{\rdisk}{r_{\mathrm{disk}}}
\nc{\mc}{M_{\mathrm{c}}}

\begin{document}
	
\title{Primordial Black Hole Interpretation in Subsolar Mass Gravitational Wave Candidate SSM200308}
	

\author{Chen Yuan}
\email{chenyuan@tecnico.ulisboa.pt}
\affiliation{CENTRA, Departamento de Física, Instituto Superior Técnico – IST, Universidade de Lisboa – UL, Avenida Rovisco Pais 1, 1049–001 Lisboa, Portugal}
\author{Qing-Guo Huang}
\email{corresponding author: huangqg@itp.ac.cn}
\affiliation{CAS Key Laboratory of Theoretical Physics,
	Institute of Theoretical Physics, Chinese Academy of Sciences,
	Beijing 100190, China}
\affiliation{School of Physical Sciences,
	University of Chinese Academy of Sciences,
	No. 19A Yuquan Road, Beijing 100049, China}
\affiliation{School of Fundamental Physics and Mathematical Sciences
Hangzhou Institute for Advanced Study, UCAS, Hangzhou 310024, China}
	
\date{\today}

\begin{abstract}
In the recent second part of the third observation run by the LIGO-Virgo-KAGRA collaboration, a candidate with sub-solar mass components was reported, which we labelled as SSM200308. This study investigates the premise that primordial black holes (PBHs), arising from Gaussian perturbation collapses, could explain SSM200308. Through Bayesian analysis, we obtain the primordial curvature power spectrum that leads to the merger rate of PBHs aligning with observational data as long as they constitute {$f_{\mathrm{PBH}}=5.66^{+58.68}_{-5.44}\times 10^{-2}$ } of the dark matter. However, while the gravitational wave (GW) background from binary PBH mergers is within current observational limits, the scalar-induced GWs associated with PBH formation exceed the constraints imposed by pulsar timing arrays, challenging the Gaussian perturbation collapse PBH model as the source of SSM200308.
\end{abstract}

\maketitle

\maketitle

\section{Introduction}
Since the first detection of a gravitational wave (GW) event by LIGO \cite{Abbott:2016blz}, which marked the dawn of GW astronomy, the \lvc\ collaboration has reported nearly a hundred GW events over the subsequent years.\cite{LIGOScientific:2016dsl,LIGOScientific:2018mvr,LIGOScientific:2020ibl,LIGOScientific:2021usb,KAGRA:2021vkt}. Prior investigations into compact binary coalescences of sub-solar mass (SSM) objects within the collected data yielded no conclusive evidence of such events \cite{LIGOScientific:2018glc,LIGOScientific:2019kan,Nitz:2021mzz,LIGOScientific:2021job,LIGOScientific:2022hai,Nitz:2022ltl,Wang:2024vfv}. However, it was not until recently, in the second part of the third observing run \cite{KAGRA:2021vkt}, that three SSM candidates were reported \cite{LIGOScientific:2022hai} and they are also analyzed in \cite{Morras:2023jvb,Prunier:2023cyv}.
These three candidates are not claimed as GW events, owing to their relatively large false alarm rates and relatively low signal-to-noise ratio (SNR). Notably, one of the candidates, SSM200308, exhibited a false alarm rate of approximately $0.2$ per year and an $\mathrm{SNR}=8.90$, positioning it on the cusp of claiming a GW detection. With ongoing enhancements in \lvc's sensitivity and the accumulation of observational data, the investigation into the nature and origins of SSM candidates is emerging as a promising and compelling field of study. Recent research has proposed primordial black holes (PBH) as a potential source of SSM candidates \cite{Yamamoto:2023tsr,Carr:2023tpt,Crescimbeni:2024cwh}, suggesting a novel avenue for understanding these enigmatic phenomena.

PBHs are formed in the very early Universe, originating from the collapse of overmassive regions \cite{Zeldovich:1967lct,Hawking:1971ei,Carr:1974nx,Carr:1975qj}. The overabundance of mass in these regions is generated by large curvature perturbations that are enhanced on scales (e.g., \cite{Pi:2017gih,Cai:2018tuh,Cotner:2016cvr,Espinosa:2017sgp,Palma:2020ejf,Pi:2021dft,Meng:2022low}) much smaller than those observed in the cosmic microwave background \cite{Planck:2018vyg}. PBHs are not only promising dark matter (DM) candidates but if a few-thousandth of DM consists of PBHs, they can also explain the GW events observed by the \lvc\ collaboration \cite{Sasaki:2016jop,Chen:2018czv,Raidal:2018bbj,DeLuca:2020qqa,Hall:2020daa,Bhagwat:2020bzh,Hutsi:2020sol,Wong:2020yig,DeLuca:2021wjr,Bavera:2021wmw,Franciolini:2021tla,Chen:2021nxo,Chen:2024dxh}.

During the formation of PBHs, second-order tensor modes will be inevitably sourced by the quadratic terms of linear curvature perturbations, known as scalar-induced gravitational waves (SIGWs) \cite{tomita1967non,Matarrese:1992rp,Matarrese:1993zf,Ananda:2006af,Baumann:2007zm,Saito:2008jc,Saito:2009jt}. The unique and model-independent scaling in the low-frequency region of the SIGW spectrum \cite{Yuan:2019wwo,Yuan:2023ofl} makes SIGW a useful tool in hunting PBHs. We refer the readers to \cite{Yuan:2021qgz,Domenech:2021ztg} for review of SIGWs.

In this paper, we explore the possibility of whether SSM200308 can be explained by PBHs generated by the collapse of Gaussian perturbations in the early Universe. We calculate the merger rate density derived from the primordial curvature power spectrum. By performing Bayesian parameter estimation using the SSM200308 data, we analyze the curvature power spectrum and examine its astrophysical implications through SIGWs and stochastic gravitational wave background (SGWB) associated with binary PBH coalescences, finally contrasting the results with current pulsar timing array (PTA) data to investigate the potential connections between PBHs and SSM200308.

\section{Bayesian analysis of SSM200308}
In this section, we perform a Bayesian inference \cite{LIGOScientific:2016kwr,LIGOScientific:2016ebi,TheLIGOScientific:2016pea,Wysocki:2018mpo,Fishbach:2018edt,Mandel:2018mve,Thrane:2018qnx} to obtain the parameters of the primordial curvature power spectrum that lead to the PBH mass function for interpreting SSM200308. Firstly, the merger rate density of PBHs for general PBH mass functions, $P(m)$, in the units of $\mathrm{Gpc}^{-3}\mathrm{yr}^{-1}$ are given by \cite{Chen:2018czv}
{
\begin{equation}\label{R12}
\begin{aligned}
\mathcal{R}_{12}(t \mid \!\vec{\theta})& =\frac{1.6 \times 10^{6}}{\mathrm{Gpc}^{3} \mathrm{yr}} f_{\mathrm{PBH}}^{\frac{53}{37}}\left(\frac{t(z)}{t_{0}}\right)^{-\frac{34}{37}} \eta^{-\frac{34}{37}}\left(\frac{M}{M_{\odot}}\right)^{-\frac{32}{37}} \\
& \times S\left[M, f_{\mathrm{PBH}}, P(m), z\right] P\left(m_{1}\right) P\left(m_{2}\right),
\end{aligned}
\end{equation}
where $M=m_1+m_2$, $\eta=m_1 m_2/M^2$}, $t_0$ is the age of the Universe, $\fpbh \equiv \Omega_{\mathrm{PBH}}/\Omega_{\mathrm{CDM}}$ represents the energy fraction of CDM in the form of PBHs and $\sigma_{\mathrm{eq}}\approx 0.005$ \cite{Ali-Haimoud:2017rtz,Chen:2018czv} stands for the variance of density perturbations of the rest CDM at matter-radiation equality. $\vth$ represents a population of parameters in the PBH mass function. The mass function is normalized as $\int P(m)\mathrm{d}m=1$. {The expression of  suppression factor, $S$, can be found in \cite{Hutsi:2020sol,Franciolini:2022tfm}. }
{We also considered the merger rate contributed by the three-body configurations, which is expressed as \cite{Vaskonen:2019jpv,Raidal:2024bmm}
\begin{equation}
    \begin{aligned}
\mathcal{R}_{12,3}(t\mid\! \vec{\theta}) & \approx \frac{7.9 \times 10^{4}}{\mathrm{Gpc}^{3} \mathrm{yr}}\left(\frac{t}{t_{0}}\right)^{\frac{\gamma}{7}-1} f_{\mathrm{PBH}}^{\frac{144 \gamma}{259}+\frac{47}{37}} \\
\times & {\left[\frac{\langle m\rangle}{M_{\odot}}\right]^{\frac{5 \gamma-32}{37}}\left(\frac{M}{2\langle m\rangle}\right)^{\frac{179\gamma }{259}-\frac{2122}{333}}(4 \eta)^{-\frac{3 \gamma}{7}-1} } \\
& \times \mathcal{K} \frac{e^{-3.2(\gamma-1)} \gamma}{28 / 9-\gamma} \overline{\mathcal{F}}\left(m_{1}, m_{2}\right) P\left(m_{1}\right) P\left(m_{2}\right),
\end{aligned}
\end{equation}
where $\gamma$ and $\mathcal{K}$ stand for the dimensionless distribution of angular momenta and the hardening of the PBH binary in encounters with other PBHs respectively. We use $\gamma=1$ and $\mathcal{K}=1$ suggested by the numeric simulation \cite{Raidal:2018bbj}. The $\overline{\mathcal{F}}$ factor is given by
\begin{equation}
    \begin{aligned}
\overline{\mathcal{F}}\left(m_{1}, m_{2}\right) & \equiv \int_{m \leq m_{1}, m_{2}} \mathrm{~d}m~ P(m) \frac{\langle m\rangle}{m} \\
& \times\left[2 \mathcal{F}\left(m_{1}, m_{2}, m\right)+\mathcal{F}\left(m, m_{1}, m_{2}\right)\right],
\end{aligned}
\end{equation}
with 
\begin{equation}
    \begin{aligned}
\mathcal{F}\left(m_{1}, m_{2}, m_{3}\right) & =m_{1}^{\frac{5}{3}} m_{2}^{\frac{5}{3}} m_{3}^{\frac{7}{9}}\left(\frac{m_{1}+m_{2}}{2}\right)^{\frac{4}{9}} \\
& \times\left(\frac{m_{1}+m_{2}+m_{3}}{3}\right)^{\frac{2}{9}}\langle m\rangle^{-\frac{43}{9}}.
\end{aligned}
\end{equation}
}

The PBH mass fraction can be obtained using threshold statistics by integrating the peaked non-linear compaction, defined as the mass excess compared to the background value within a given radius \cite{Harada:2015yda}, and the result is
\begin{equation}\label{beta}
\beta=\int_{\mathcal{C}_c}^{\infty} \mathrm{d} \mathcal{C}\left(r_m\right) \int_{-\infty}^0 \mathrm{~d} \mathcal{C}^{\prime \prime}\left(r_m\right)\frac{m}{M_H} \mathcal{P}\left[\mathcal{C}\left(r_m\right), \mathcal{C}^{\prime \prime}\left(r_m\right)\right],
\end{equation}
where $\mathcal{C}(r)=\mathcal{C}_g-\frac{1}{4\Phi}\mathcal{C}_g^2$ and $\mathcal{C}_g=-2\Phi r \zeta'(r)$ is linearly related to the comoving curvature perturbation. Here, $\Phi \equiv 3(1+w)/(5+3w)$ and $w$ is the equation of state. {In this paper, we consider the softening of the equation of state during the QCD phase transition \cite{Musco:2012au,Saikawa:2018rcs} where the threshold value of forming PBHs will slightly decrease \cite{Borsanyi:2016ksw,Saikawa:2018rcs,Franciolini:2022tfm,Musco:2023dak} since there is less pressure to resist the collapse overdensities.}
$r_m$ in Eq.~(\ref{beta}) represents the distance between the location where $\mathcal{C}_g$ gains its maximum from and the origin of the peak in curvature perturbation, $\zeta(0)$. The mass of PBH over the horizon mass is given by $m/M_H =\kappa\left(\mathcal{C}(r_m)-\mathcal{C}_c\right)^\gamma$ \cite{Choptuik:1992jv,Evans:1994pj,Niemeyer:1997mt}. {For simplicity, the scaling parameters $\gamma$ and $\kappa$ during the QCD transition are set to $\kappa=3.3$ and $\gamma=0.36$ \cite{Koike:1995jm}.}
The second-order derivative of the compact function is given by
\begin{equation}
\mathcal{C}^{\prime \prime}\left(r_m\right)=\mathcal{C}_g^{\prime \prime}\left(r_m\right)\left[1-\frac{1}{2\Phi} \mathcal{C}_g\left(r_m\right)\right].
\end{equation}
Throughout this paper, we consider type I PBHs, namely PBHs generated in the regime $\mathcal{C}_g<2\Phi$. For type I PBHs, the local maximum of the non-linear compact function coincides with the linear one. We only consider Gaussian curvature perturbation so that the probability density function of  $\mathcal{C}_g$ and $\mathcal{C}''_g$ follow joint Gaussian distribution \cite{Ferrante:2022mui,Gow:2022jfb,Ianniccari:2024bkh}:
\begin{equation}
\begin{aligned}
\mathcal{P}\left[-\frac{1}{4} r_m^2 \mathcal{C}_g^{\prime \prime}\left(r_m\right), \mathcal{C}_g\left(r_m\right)\right] & =\frac{\exp \left(-\vec{V}^T \Sigma^{-1} \vec{V} / 2\right)}{2 \pi \sqrt{\operatorname{det} \Sigma}} , \\
\vec{V}^T & =\left[-\frac{1}{4} r_m^2 \mathcal{C}_g^{\prime \prime}\left(r_m\right), \mathcal{C}_g\left(r_m\right)\right], \\
\Sigma & =\left(\begin{array}{cc}
\sigma_2^2 & \sigma_1^2 \\
\sigma_1^2 & \sigma_0^2
\end{array}\right) .
\end{aligned}
\end{equation}
To calculate $\beta$, we transform the variables in the integrand in Eq.~(\ref{beta}) to $\mathcal{C}_g$ and $\mathcal{C}''_g$, and then integrate over the new probability density function of the joint Gaussian distribution (also with the determinant of the Jacobian matrix) in the new range of integration.

The correlations can be calculated in Fourier space such that \cite{Ferrante:2022mui}
\begin{equation}
\sigma_j^2=\int \frac{d k}{k} W^2(k, R) T^2(k, R) P_\zeta(k) k^{2 j},
\end{equation}
where $P_\zeta$ represents the primordial curvature power spectrum for which we adopt a widely used model \cite{Yuan:2019wwo,Pi:2020otn,Domenech:2020kqm,Yuan:2020iwf,Yuan:2021qgz,Meng:2022ixx,Yuan:2023ofl,Gorji:2023ziy,Franciolini:2023pbf,Papanikolaou:2024kjb}:
\begin{equation}
P_{\zeta}(k)=\frac{A}{\sqrt{2 \pi} \sigma_*} \exp \left(-\frac{\ln ^2\left(k / k_{\star}\right)}{2 \sigma_*^2}\right),
\end{equation}
where the amplitude, peak location and broadness are controlled by $A$, $k_*$ and $\sigma_*$ respectively.
The window function $W(k,R)=\exp(-x^2/4)$ with $x\equiv k R$ is chosen to pick out relevant perturbations in PBH formation \cite{Ando:2018qdb,Young:2019osy}. The transfer function is obtained assuming radiation dominant, namely
\begin{equation}
T(k, R)=3\left[\frac{\sin (k R / \sqrt{3})-(k R / \sqrt{3}) \cos (k R / \sqrt{3})}{(k R / \sqrt{3})^3}\right]
\end{equation}
Finally, the mass function of PBHs is given by 
\begin{equation}
	\fpbh(m) = \fpbh P(m) m=\frac{1}{\Omega_{\mathrm{CDM}}}\left(\frac{M_{\mathrm{eq}}}{m}\right)^{1 / 2} \beta\left(m\right),
\end{equation}
where the horizon mass at matter-radiation equality is $M_{\mathrm{eq}}\simeq 2.8\times10^{17}\Msun$.
Compared to $P(m)$, $\fpbh(m)$ is the more commonly used mass function in observations and it is normalized as
\begin{equation}
    \int\fpbh(m)\mathrm{d} \log m = \fpbh.
\end{equation}

Given the data of SSM200308, the likelihood for an inhomogeneous Poisson process reads \cite{Wysocki:2018mpo,Fishbach:2018edt,Mandel:2018mve,Thrane:2018qnx}
\begin{equation}\label{likelihood}
    p(data|\vth, R) \propto e^{-\beta(\vth)} \int \rd\vla\ p(\vla|data) \ \mR_{12}(\vla|\vth),
\end{equation}
with $\vla \equiv \{m_1, m_2,z\}$, $p(\vla|data)$ is the posterior of SSM200308 obtained in \cite{Prunier:2023cyv} and $\beta(\vth) \equiv \int \rd\vla\ VT(\vla)\ \mR_{12}(\vla|\vth)$ where $VT(\vla)$ is the spacetime sensitivity volume of \lvc\ detectors \cite{LIGOScientific:2022hai}. 

The parameters in the PBH mass function are $\{A, k_*, \sigma_*\}$ and the priors that are used in the analysis are uniformly distributed in $\log_{10}A\in[-2,0]$, $\log_{10} k_*\in[6.5,8.0]$ and $\log_{10} \sigma_*\in[-1,0]$. Note that the value of $\fpbh$ can be obtained once $\{A, k_*, \sigma_*\}$ is fixed. 


\begin{figure}
    \centering
    \includegraphics[width=0.9\columnwidth]{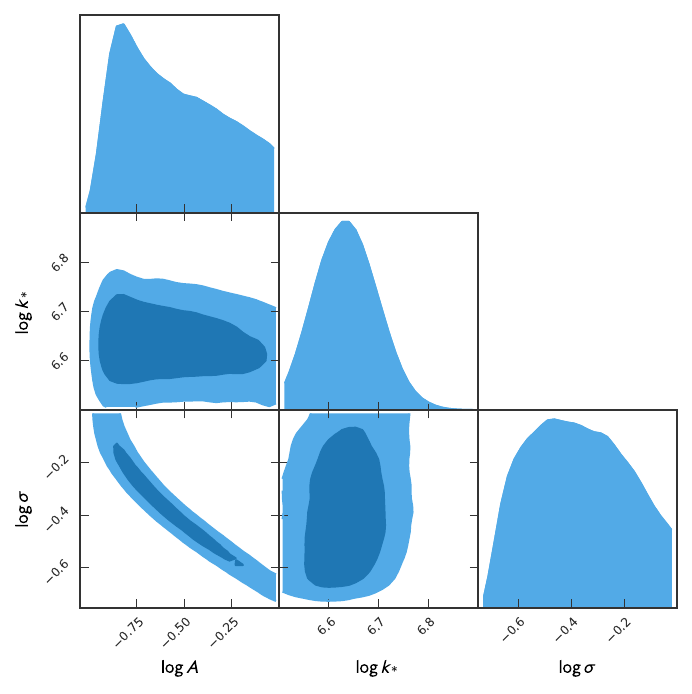}
    \caption{The posterior distribution of the Bayesian parameter estimation. The contours in the 2D distribution represent $1\sigma$ and $2\sigma$ regions respectively.}
    \label{post}
\end{figure}

We report our analysis in Fig.~\ref{post}. The median value of the parameters and their $95\%$ equal-tailed credible intervals are {$A=0.26^{+0.56}_{-0.13}$, $k_*=4.31^{+1.16}_{-0.84}\times 10^6\mathrm{Mpc}^{-1}$ and $\sigma=0.42^{+0.44}_{-0.19}$ }. Moreover, the fraction of PBHs in DM is evaluated to be {$\fpbh=5.66^{+58.68}_{-5.44}\times 10^{-2}$ } and the local merger rate is given by { $\mathcal{R}=3953^{+20939}_{-3371}\mathrm{Gpc}^{-3}\mathrm{yr}^{-1}$}. The sensitivity volume values from the O3 run in the SSM search for SSM200308 is $\left\langle VT \right\rangle\simeq 3\times 10^{-4}\mathrm{Gpc}^{3}\mathrm{yr}$ \cite{LIGOScientific:2022hai}. Therefore, the expected event numbers are { $N=\mathcal{R}\left\langle VT \right\rangle=1.09^{+5.76}_{-0.93}$} which is consistent with \cite{Phukon:2021cus,LIGOScientific:2022hai}.

\begin{figure}
    \centering
    \includegraphics[width=0.9\columnwidth]{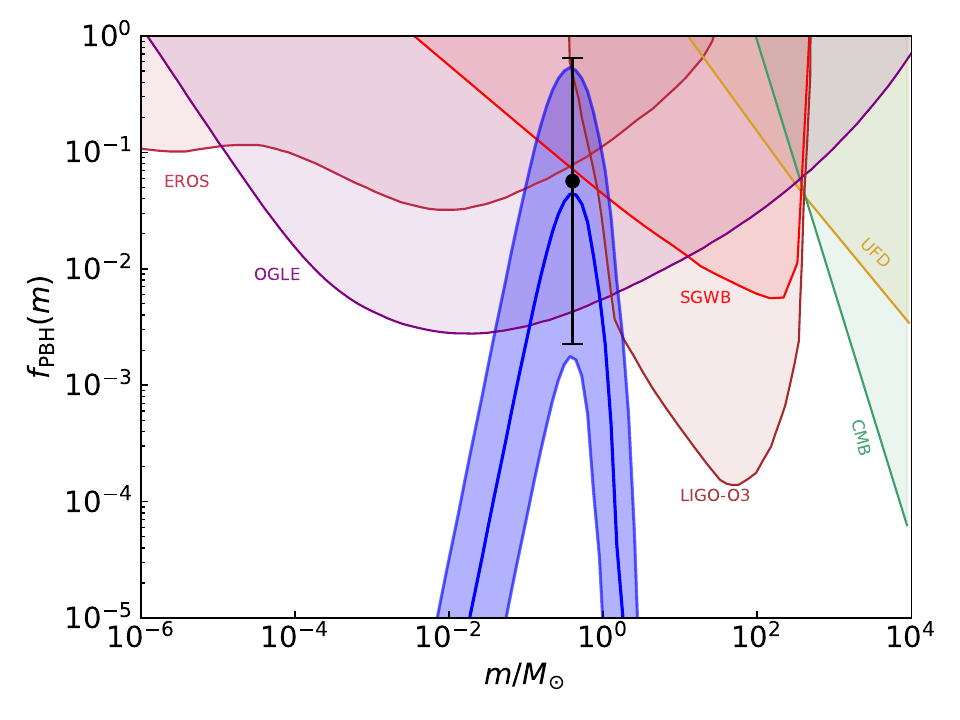}
    \caption{The posterior distribution of the PBH mass function {and current constraints on $\fpbh$}. The blue bands represent PBH mass function at $95\%$ confidence level. {The error bar stands for $\fpbh=5.66^{+58.68}_{-5.44}\times 10^{-2}$ }.}
    \label{massfunc}
\end{figure}

Furthermore, we demonstrate the PBH mass function {and $\fpbh$} in Fig.~\ref{massfunc}, together with the median mass of SSM200803. We also show the constraints on $\fpbh$ by EROS/MACHO microlensing at $95\%$ CL \cite{EROS-2:2006ryy}, OGLE microlensing at $95\%$ CL \cite{Mroz:2024mse,Mroz:2024wag}, {SGWB from binary PBHs by LIGO-Virgo-KAGRA O3 \cite{Nitz:2022ltl} at $90\%$ CL and the LIGO-O3 results in \cite{Hutsi:2020sol} at $2\sigma$ CL by assuming all the O3 events have astrophysical origin}, dynamical heating of ultra-faint dwarf galaxies (UFD) \cite{Brandt:2016aco} and accretion constraints of CMB at $95\%$ \cite{Ali-Haimoud:2016mbv,Blum:2016cjs,Horowitz:2016lib,Chen:2016pud,Poulin:2017bwe}. It can be seen from Fig.~\ref{massfunc} that the result of current constraints has some overlap with $\fpbh$. As future data accumulate, it may be possible to verify or falsify these SSM PBHs. {We remind the readers that all these constraints on $\fpbh$ assume a monochromatic PBH mass function and a precise comparison should generalize these constraints to an extended mass function scenario (see e.g., \cite{Carr:2017jsz}). However, the data is in favour of a small width power spectrum, therefore the resulting mass function is approximately monochromatic and the result of $\fpbh$ in Fig.~\ref{massfunc} would give readers a rough order of magnitude comparison.}

\section{SGWB associated with PBHs}
The incoherent superposition of the GWs from all the emitting sources all over the sky will generate an SGWB. In this section, we will demonstrate the SIGWs associated with the formation of PBHs and the SGWB from the merger of binary PBHs if the SSM200308 can be explained by PBHs.

\subsection{Scalar-Induced Gravitatiaonal Waves}
The second-order tensor modes in the perturbed FRW metric will be sourced by the quadratic terms of linear curvature perturbations. These second-order GWs are inevitably generated during the formation of PBHs and their unique scaling in the low-frequency region provides a smoking gun in hunting PBHs \cite{Yuan:2019wwo,Yuan:2023ofl}. The perturbed metric in Newtonian gauge in the absence of vector perturbations, linear tensor perturbations and anisotropies stress is given by
\begin{equation}
\mathrm{d} s^2=a^2\left\{-(1+2 \phi) \mathrm{d} \eta^2+\left[(1-2 \phi) \delta_{i j}+\frac{h_{i j}}{2}\right] \mathrm{d} x^i \mathrm{~d} x^j\right\},
\end{equation}
where the Bardeen potential $\phi$ is related to the comoving curvature perturbations through $\phi=-2/3\zeta$. The equation of motion for the second-order tensor mode is given by
\begin{equation}\label{eomh}
	h_{i j}^{\prime \prime}+2 \mathcal{H} h_{i j}^{\prime}-\nabla^{2} h_{i j}=-4 \mathcal{T}_{i j}^{\ell m} \mathcal{S}_{\ell m},
\end{equation}
where the projection operator $\mathcal{T}_{i j}^{\ell m}$ picks out the transverse and traceless part of the source term. During radiation dominant, the source term is given by
\begin{equation}
	S_{i j}=2 \phi \partial_{i} \partial_{j} \phi-\frac{4}{3(1+w)}\left(\partial_{i} \phi+\frac{\partial_{i} \phi^{\prime}}{\mathcal{H}(\eta)}\right)\left(\partial_{j} \phi+\frac{\partial_{j} \phi^{\prime}}{\mathcal{H}(\eta)}\right).
\end{equation}
Following \cite{Kohri:2018awv}, the solution to Eq.~(\ref{eomh}) can be solved by the Green's function and the energy spectrum of SIGWs, $\Omega_{\mathrm{GW}}$, which is defined as the energy of GWs per logarithm wavelength normalized by the critical energy of the Universe, can be computed in a semi-analytical way such that \cite{Kohri:2018awv}
\begin{equation}
    \begin{aligned}
        	\Omega_{\mathrm{GW}}(f)&=\frac{\Omega_{\mathrm{r}}}{6}\left(\frac{g_*}{g_*^0}\right)\left(\frac{g_{*s}}{g_{*s}^0}\right)^{-4/3} \int_{0}^{\infty} \mathrm{d} u \int_{|1-u|}^{1+u} \mathrm{~d} v \\
	&	\frac{v^{2}}{u^{2}}\left[1-\left(\frac{1+v^{2}-u^{2}}{2 v}\right)^{2}\right]^{2} \mathcal{P}_{\zeta}(u k) \mathcal{P}_{\zeta}(v k) \overline{I^{2}(u, v)},
    \end{aligned}
\end{equation}
where $k=2\pi f$ and $\Omega_r=9\times10^{-5}$ is the density parameter for radiation. $g_*$ and $g_{*s}$ stand for the effective degrees of freedom of entropy and energy respectively. The kernel function $\overline{I^{2}(u, v)}$ is given by \cite{Kohri:2018awv}:
\begin{equation}
    \begin{aligned}
&\overline{I^2(u,v)}=\frac{9(u^2+v^2-3)^2}{32u^6v^6}\Bigg\{\Big(-4uv+(u^2+v^2-3)\\
&\times \ln\Big|{3-(u+v)^2\over3-(u-v)^2}\Big|\Big)^2+\pi^2\left(u^2+v^2-3\right)^2\Theta(u+v-\sqrt{3})\Bigg\}.
    \end{aligned}
\end{equation}

\subsection{SGWB form binary PBH coalescences}
The SGWB generated by binary PBH coalescences all over the sky can be evaluated as \cite{Phinney:2001di,Regimbau:2008nj,Zhu:2011bd,Zhu:2012xw}
\begin{equation}\label{SGWB}
    \Omega_{\mathrm{GW}}(f)=\frac{f}{\rho_c} \int d M_1d M_2 d z \frac{d t}{d z} R(z,M_1,M_2)\frac{d E_s}{d f_s},
\end{equation}
with $dt/dz$ being the derivative of the lookback time to the redshift, which takes the form
\begin{equation}
    \frac{d t}{d z}=\frac{1}{H_{0} \sqrt{\Delta}(1+z)}.
\end{equation}
Here $H_0$ is the Hubble value by today, $\Delta = \Omega_{r}(1+z)^{4}+\Omega_{m}(1+z)^{3}+\Omega_{\Lambda}$, $\Omega_m$ and $\Omega_\Lambda$ are density parameters for matter and dark energy respectively \cite{Planck:2018vyg}. The energy spectrum of a single PBH coalescence event in the source frame, ${d E_s}/{d f_s}$, can be found in \cite{Cutler:1993vq,Chernoff:1993th,Zhu:2011bd}. Note that Eq.~(\ref{SGWB}) also includes the contribution of resolvable events. However, as we will see, the result of SGWB is far below the current upper limits even if it is overestimated.

\begin{figure}
    \centering
    \includegraphics[width=0.9\columnwidth]{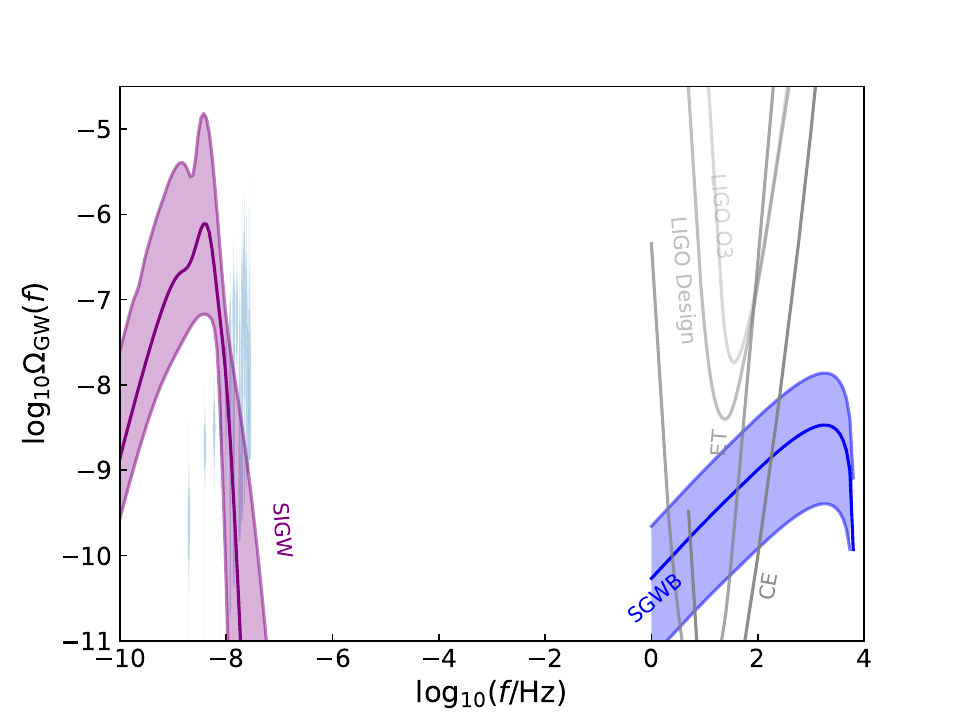}
    \caption{The SIGW (purple lines) and SGWB from binary PBHs (blue lines) if SSM200803 can be explained by PBHs. The violin polot represents the NANOGrav 15yr dataset \cite{NANOGrav:2023gor}. The purple and blue bands represent the region at $95\%$ confidence level for SIGWs and SGWB from binary PBHs respectively.}
    \label{Ogw}
\end{figure}

The median value and the $95\%$ intervals of SIGWs and SGWB from binary PBH coalescences are illustrated in Fig.~\ref{Ogw} using the posterior of $A$, $K_*$ and $\sigma$ obtained in Section. II together with the power-law integrated sensitivity curves of LIGO O3, LIGO at design sensitivity \cite{Thrane:2013oya}, Cosmic Explorer (CE) \cite{Evans:2016mbw}, Einstein Telescope (ET) \cite{Punturo:2010zz} and the constraints from NANOGrav 15yr dataset \cite{NANOGrav:2023hvm}.

Although we do not subtract resolvable events, the SGWB still does not conflict with the existing upper limits, it can be seen that the SGWB form binary PBHs is far below the sensitivity of LIGO O3 and it could be verified by CE and ET in the future.  However, the resulting SIGWs exceed the upper limits of NANOGrav 15yr, indicating a discrepancy for PBHs generated by the collapse of Gaussian perturbations to explain SSM200803.

\section{Conclusion and Discussion}
In this study, we have delved into the potential explanation of SSM200308 as a consequence of PBHs emerging from the collapse of Gaussian perturbations in the early universe. Through a comprehensive Bayesian parameter estimation, we obtain the corresponding primordial curvature power spectrum that generates the PBHs for explaining SSM200308. Our findings indicate that although the merger rates of PBHs could be reconciled with the data and the SGWB from binary PBHs is consistent with current observations as long as PBHs constitute a few thousandths of the DM, the accompanied SIGWs during the formation of PBHs present a significant challenge. The SIGWs generated by the primordial curvature power spectrum would exceed the limits set by PTA observations, suggesting a discrepancy that questions the overdensities collapse as the origin of PBHs responsible for SSM200308.

However, SIGWs are related to PBHs formed by the collapse of overdensities. Given that the merger rate of PBHs is not sensitive with $P(m)$ and it is mainly dependent on $\fpbh$. Therefore, the SGWB from binary PBH mergers would also be consistent with existing observations if the PBHs have other formation mechanisms, such as first-order phase transition \cite{LISACosmologyWorkingGroup:2023njw}. As a result, the SIGWs would disappear if PBHs had other formation mechanisms. In that case, our results cannot entirely rule out the possibility of PBHs in explaining SSM200803. Moreover, the PBH formation model contains uncertainties in calculating $\fpbh$ from the curvature power spectrum. For instance, adopting the peaks theory \cite{Bardeen:1985tr} to evaluate the PBH abundance leads to smaller SIGW \cite{Franciolini:2022tfm} and thus alleviates the tension with PTAs.

It's worth mentioning that if primordial non-Gaussian effects are present during the formation of PBHs, the amplitude of the SIGWs could be suppressed \cite{Nakama:2016gzw,Garcia-Bellido:2017aan,Unal:2018yaa,Cai:2018dig,Cai:2019amo,Yuan:2020iwf,Ragavendra:2020sop,Adshead:2021hnm,Abe:2022xur,Yuan:2023ofl,Li:2023xtl,Li:2024zwx,Perna:2024ehx}, potentially explaining the results from NANOGrav \cite{Liu:2023ymk,Franciolini:2023pbf}. Moreover, in this work, we focused on using the data from SSM200803. An even more intriguing result would be to use both results from NANOGrav 15yr and SSM200803 to place constraints on PBHs and the primordial non-Gaussianities. We leave this part for future work.

\vspace{5mm}

{\it Acknowledgments. }
We thank Prunier {\it et al.} for sharing the data on the posterior distribution of SSM200803 \cite{Prunier:2023cyv}.  C.Y. thanks Antonio Riotto, Zu-cheng Chen, Yu-mei Wu and Shi Pi for useful discussions and valuable advice. This work is supported by the National Key Research and Development Program of China Grant No.2020YFC2201502, grants from NSFC (grant No. 11991052, 12047503), Key Research Program of Frontier Sciences, CAS, Grant NO. ZDBS-LY-7009. 

This project has received funding from the European Union's Horizon 2020 research and innovation programme under the Marie Sklodowska-Curie grant agreements No 101007855 and No 101131233.

\bibliography{./ref}
 
\end{document}